\documentclass[prd,amsmath,amssymb,superscriptaddress,nofootinbib]{revtex4}

\usepackage{graphicx}
\usepackage{dcolumn}
\usepackage{bm}

\normalbaselines \topmargin -0.25 truein \oddsidemargin 0.30
truein \evensidemargin 0.30 truein 

\newcommand{\D}{\hat{D}}

\begin{document}

\title{Decay of a multi-axionic SU(N) symmetric color aether in the early Universe as an origin of emergence of a many-component dark matter}

\author{Alexander B. Balakin}
\email{Alexander.Balakin@kpfu.ru} \affiliation{Department of General
Relativity and Gravitation, Institute of Physics, Kazan Federal University, Kremlevskaya
str. 18, Kazan 420008, Russia}
\author{Gleb B. Kiselev}
\email{kiselev.gleb.97@gmail.com} \affiliation{Department of General
Relativity and Gravitation, Institute of Physics, Kazan Federal University, Kremlevskaya
str. 18, Kazan 420008, Russia}

\date{\today}

\begin{abstract}
We establish a new SU(N) symmetric model of interaction between the fields of four types: the multiplet of vector fields, which describes the so-called color aether,  the multiplet of pseudoscalar fields, which is associated with the multi-component cosmic dark matter, the gauge and gravitational fields. The extended Lagrangian of the model contains a new constructive element, which is based on the covariant SU(N) symmetric divergence of the multiplet of vector fields; this new element, being the multiplet of scalars from the point of view of spacetime transformations and the color vector from the point of view of the SU(N) group space, gives us the possibility to formulate properly the multi-axionic extension of the Peccei-Quinn theory. The hypothesis of a spontaneous polarization of the multi-axionic color aether in the early Universe is presented. The set of self-consistent master equations of the model is derived. An application to cosmology is considered: the obtained master equations are solved for the truncated test model based on the Bianchi-I spacetime platform.
\end{abstract}

\maketitle



\section{Introduction}\label{Intro}

The logic of deepening  our knowledge of the physics of the early Universe suggests that many fundamental fields, which are known today and were experimentally studied during last century, could have had their own SU(N) symmetric prototypes, the evolution of which was regulated by the corresponding gauge fields connecting them \cite{Rubakov1,Rubakov2}.

For instance, there is a consensus among modern cosmologists that the dark matter, as one of the main structural elements of any cosmological model, is multi-component, and one of its fractions is, highly likely, formed by axions, massive pseudo-Goldstone bosons \cite{PQ}-\cite{Marsh}.
It is quite logical to assume that in the early era of the Universe expansion there existed a SU(N) symmetric multiplet of pseudoscalar fields $\phi^{(a)}$, one of which has now turned into the relic light axions that form cold dark matter and is described by the pseudoscalar singlet $\phi$, while the others could have made up the so-called warm and hot fractions of dark matter. The symbol $(a)$ denotes the  group index in the adjoint representation of the SU(N) group, and this Latin index runs over the values $\left\{(1), (2), ..(N^2{-}1)\right\}$.

Similarly, now there exists the model indicated as dynamic aether \cite{J1} - \cite{J10}, which was introduced as a version of the vector-tensor branch of the science entitled Modified Theories of Gravity (see, e.g., \cite{O1,Bamba2,O2,03}).
This theory operates with a timelike unit four-vector field $U^j$, which can be interpreted as the velocity four-vector of the cosmic substratum indicated as the dynamic aether, one of the candidates for the role of a dark energy. It is quite logical to assume that in the early Universe there existed a SU(N) symmetric multiplet of vector fields $U^{j (a)}$ \cite{color1,color2,color3}, which in a later era degenerated into a singlet $U^j$, representing the four-vector of the velocity of the dynamic aether.

In order to maintain SU(N) invariance of the dynamic theory of evolution of the pseudoscalar and vector multiplets, one needs to include into the Lagrangian the extended covariant derivatives of these fields \cite{Akhiezer}, thus, the multiplet of potential co-vectors $A^{(a)}_m$, which describes the SU(N) symmetric gauge field,  appears inevitably as the structural element of the theory. Let us emphasize that we do not include spinor fields, associated with quarks, nucleons, etc.,  into the model under consideration, thus providing the gauge fields to be associated with the color aether and multi-axion configuration, exclusively.

The goal of this work is to establish the full-format self-consistent formalism of evolution of the SU(N) symmetric system, which contains three interacting multiplets $\left\{\phi^{(a)} \right\}$,  $\left\{U^{j(a)} \right\}$ and $\left\{A^{(a)}_m \right\}$ and the gravity field. In other words, we consider the extension of the U(1) symmetric Einstein - Maxwell - aether - axion theory \cite{EMAA1,EMAA2,EMAA3,BS} and establish the SU(N) symmetric model, which, in principle, could be indicated as Einstein - Yang - Mills - color aether - multi-axion theory.

The key element and the starting point of our modeling is the formulation of the terms in the total Lagrangian, which describe the extension of the Peccei-Quinn term $\frac14 \phi F^{*(a)}_{mn} F^{(b)mn} G_{(a)(b)}$ with the pseudoscalar (axion) field $\phi$ in front (see \cite{PQ}) to the case with the pseudoscalar multiplet $\left\{\phi^{( c)} \right\}$ instead of the axionic singlet. The usage of two elements only: of the product of the Yang-Mills field strength $F^{(b)mn}$ with its dual pseudotensor $F^{*(a)}_{mn}$, and of the metric in the group space $G_{(a)(b)}$, cannot provide the appropriate inclusion of the multiplier $\phi^{(c)}$, which adds supplementary group index. The formal problem appearing in this context is solved in our work by using the new constructive element, the extended covariant divergence of the vector fields $\hat{D}_m U^{(a) m} = \Omega^{(a)}$, which gives us an additional SU(N) symmetric multiplet of scalars $\left\{\Omega^{(a)} \right\}$.

Now we consider a special ansatz, that in one of the early epochs of the Universe expansion a series of Phase Transitions took place, as a result of which  the multiplet $\left\{U^{j(a)} \right\}$ turned into the singlet $U^j$, and the singlet $\phi$ was released from the multiplet $\left\{\phi^{(a)} \right\}$. In the works \cite{color2,color3} we indicated the transition from the SU(N) symmetric  aether to the standard dynamic aether as a spontaneous polarization of the color aether. This idea was based on the analogy with Phase Transitions in ferroelectric materials under the influence of the temperature drop \cite{ferro}, since, passing the Curie point, the ferroelectric material leaves the symmetrical phase and enters the dissymmetrical one.
Similarly, it was assumed that at the moment $t_*$ of the cosmological time the SU(N) symmetric multiplet of vector fields $\left\{U^{j(a)} \right\}$  was  converted into the one-dimensional bundle of vectors parallel in the group space, i.e., $U^{j(a)} = q^{(a)} U^j$, where $q^{(a)}$ is a constant vector in the group space. Previously, the idea of the alignment of the SU(N) symmetric multiplets was successfully applied for the Yang-Mills field configurations (see, e.g., \cite{Y,G}).
The novelty of the present work is the following:  we assume, that at some moment of the cosmological time $t_{**}$ the multiplet of the pseudoscalar fields  $\left\{\phi^{(a)} \right\}$ also degenerates, i.e., $\phi^{(a)} = \tilde{Q}^{(a)}\phi$, where $\tilde{Q}^{(a)}$ is a constant vector in the group space. In this context we speak about an axionization of the multiplet of pseudoscalar fields. The third Phase Transition is associated with the alignment of the Yang-Mills field potentials in the group space, i.e., $A^{(a)}_m \to Q^{(a)} A_m$ at $t=t_{***}$ (in this context the multiplet of gauge potentials becomes quasi-Abelian). Generally, we assume that $t_{*}\neq t_{**} \neq t_{***}$ and $q^{(a)} \neq Q^{(a)} \neq \tilde{Q}^{(a)}$, thus providing a lot of versions of the corresponding stories listed below.
Clearly, the simplest story takes place, when $t_{*}=t_{**}=t_{***}$ and $q^{(a)} = Q^{(a)} = \tilde{Q}^{(a)}$; in other words, all three degeneration processes are simultaneous and self-similar. This test model is analyzed here as a first application of the developed theory. An alternative to this test case, i.e., the case for which the polarization of color aether and alignment of the Yang-Mills potentials occurred simultaneously, but the color axions remained non-Abelian for some time after this Phase Transition, is considered in the separate paper.

This paper is organized as follows. In Section II we recall the basic concept of the SU(N) symmetric field formalism, we suggest the extended Lagrangian of interacting fields,  and derive the self-consistent system of coupled master equations for the multiplet of the vector fields $\left\{U^{j(a)} \right\}$,  for the multiplet of pseudoscalar fields $\left\{\phi^{(a)} \right\}$, for the multiplet of Yang-Mills potentials $\left\{A^{(a)}_m \right\}$, and for the gravity field. In Section III  we discuss the possible mechanisms of transformation of the SU(N) symmetric color aether to the canonic  dynamic aether, which is accompanied by the transformation of the multipled of pseudoscalar field to the axionic singlet. As an application, we consider the test model, which describes a total simultaneous degeneration of vector, pseudoscalar and gauge potential multiplets in the framework of Bianchi-I cosmological model. A look at the prospects of this theory is presented in Section IV.

\section{The formalism}

We assume that the model under consideration is constructed to describe the interaction within the following quartet of fields. First, we deal with the color aether, which is presented by the multiplet of vector fields $U^{(a)j}$, where the group index $(a)$ relates to the SU(N) symmetry group and thus enumerates $N^2{-}1$ four-vectors from this multiplet. The second element of the model is the SU(N) symmetric multiplet of the Yang-Mills gauge field potentials $A^{(a)}_k$. The SU(N) symmetric multiplet of pseudoscalar  fields $\phi^{(a)}$ relates to the third element of the model. Finally, the gravitational field is the fourth element of the theory; it is described in the Einstein-Hilbert scheme using the spacetime metric $g_{mn}$.

\subsection{Basic elements of the SU(N) symmetric theory}

\subsubsection{Geometry of the group space}

As usual, the basis in the group space \cite{Rubakov1,Akhiezer} is constructed using the Hermitian traceless generators of the SU(N) group, ${\bf t}_{(a)}$, satisfying
the commutation relations
\begin{equation}
\left[ {\bf t}_{(a)} , {\bf t}_{(b)} \right] = i  f^{(c)}_{\
(a)(b)} {\bf t}_{(c)} \,. \label{fabc}
\end{equation}
Here and below $f^{(c)}_{\ (a)(b)}$ are the structure constants of the SU(N) gauge group.
The scalar product of the generators ${\bf t}_{(a)}$ and ${\bf t}_{(b)}$ is defined to introduce the metric in the group space, $G_{(a)(b)}$:
\begin{equation}
G_{(a)(b)} = \left( {\bf t}_{(a)} , {\bf t}_{(b)} \right) \equiv 2 {\rm Tr} \
{\bf t}_{(a)} {\bf t}_{(b)}    \,.
\label{scalarproduct}
\end{equation}
The structure constants with three subscripts
\begin{equation}
f_{(c)(a)(b)} \equiv G_{(c)(d)} f^{(d)}_{\ (a)(b)} =
 {-} 2 i
{\rm Tr}  \left[ {\bf t}_{(a)},{\bf t}_{(b)} \right] {\bf
t}_{(c)}   \label{fabc1}
\end{equation}
are antisymmetric with respect to transposition of any two indices.
Following the works \cite{Rubakov1,Akhiezer} we use the canonic basis ${\bf t}_{(a)}$, which provides the relationships
\begin{equation}
\frac{1}{N} f^{(d)}_{\ (a)(c)} f^{(c)}_{\ (d)(b)} =
\delta_{(a)(b)} = G_{(a)(b)} \,.   \label{ff}
\end{equation}
The structure constants satisfy the Jacobi conditions
\begin{equation}
f^{(a)}_{\ (b)(c)} f^{(c)}_{\ (e)(h)}  {+} f^{(a)}_{\ (e)(c)} f^{(c)}_{\ (h)(b)} {+} f^{(a)}_{\ (h)(c)} f^{(c)}_{\ (b)(e)} {=} 0 \,. \label{jfabc}
\end{equation}

\subsubsection{Fields in the adjoint representation}

The Yang-Mills field potential ${\bf A}_m$ and the Yang-Mills field strength ${\bf F}_{ik}$ in the fundamental representation are connected with the multiplets of reals fields $A^{(a)}_i$ and $F^{(a)}_{ik}$ appeared in the adjoint representation, as follows:
\begin{equation}
{\bf A}_m = - i g {\bf t}_{(a)} A^{(a)}_m \,, \quad {\bf F}_{mn} = - i g {\bf t}_{(a)} F^{(a)}_{mn}  \,, \label{represent}
\end{equation}
where $g$ is the coupling constant.
The multiplet of Yang-Mills tensors $F^{(a)}_{mn}$ is presented as
\begin{equation}
F^{(a)}_{mn} = \nabla_m
A^{(a)}_n - \nabla_n A^{(a)}_m + g f^{(a)}_{\ (b)(c)}
A^{(b)}_m A^{(c)}_n \,. \label{46Fmn}
\end{equation}
Here and below $\nabla_m$ is the covariant derivative.
The multiplet of dual pseudotensors $F^{*ik}_{(a)}$
\begin{equation}
F^{*(a)ik} = \frac{1}{2}\epsilon^{ikls} F_{ls}^{(a)}
\label{dual}
\end{equation}
are based on the universal Levi-Civita pseudotensor $\epsilon^{ikls}=\frac{E^{ikls}}{\sqrt{-g}}$ with $E^{0123}=1$ and $g = {\rm det}(g_{mn})$.
The multiplet of vector fields $U^{(a)m}$ appears due to the decomposition
\begin{equation}
{\bf U}^m =  {\bf t}_{(a)} U^{(a)m} \,.  \label{Urepresent}
\end{equation}
We assume that the vector fields satisfy the condition
\begin{equation}
G_{(a)(b)} U^{(a) m} U^{(b)n} g_{mn} = 1 \,,
\label{4represent}
\end{equation}
which is the direct generalization of the normalization condition $g_{mn}U^m U^n=1$ in the canonic Einstein-aether theory.
The extended (gauge covariant) derivative $\hat{D}_m$ is defined as follows:
\begin{equation}
\D_m {\cal Q}^{(a) \cdot \cdot \cdot}_{\cdot \cdot \cdot (d)} \equiv
\nabla_m {\cal Q}^{(a) \cdot \cdot \cdot}_{\cdot \cdot \cdot (d)} +
 {\cal G} f^{(a)}_{\ (b)(c)} A^{(b)}_m {\cal Q}^{(c) \cdot \cdot
\cdot}_{\cdot \cdot \cdot (d)} -
{\cal G} f^{(c)}_{\ (b)(d)}
A^{(b)}_m {\cal Q}^{(a) \cdot \cdot \cdot}_{\cdot \cdot \cdot (c)} +...
\,,
\label{D0}
\end{equation}
where ${\cal Q}^{(a) \cdot \cdot \cdot}_{\cdot \cdot \cdot (d)}$ is arbitrary tensor in the group space \cite{Akhiezer}.
For the vector fields this formula gives
\begin{equation}
\D_m U^{(a)}_n \equiv \nabla_m
U^{(a)}_n + g f^{(a)}_{\ (b)(c)} A^{(b)}_m U^{(c)}_n
\,. \label{DU}
\end{equation}
The metric $G_{(a)(b)}$ and the
structure constants $f^{(d)}_{\ (a)(c)}$ are the gauge covariant constant tensors in the group space:
\begin{equation}
\D_m G_{(a)(b)} = 0 \,, \qquad
\D_m
f^{(a)}_{\ (b)(c)} = 0 \,. \label{DfG}
\end{equation}
The dual tensor $F^{*ik}_{(a)}$, due to the definitions (\ref{dual}) and the Jacobi conditions  (\ref{jfabc}),
 satisfies the relations
\begin{equation}
\hat{D}_k F^{*ik}_{(a)} = 0 \,. \label{Aeq2}
\end{equation}
Similarly, one can define the multiplet of pseudoscalar fields and its extended derivative:
\begin{equation}
{\bf \Phi} =  {\bf t}_{(a)} \phi^{(a)} \,, \quad \D_k \phi^{(a)} = \nabla_k \phi^{(a)} + g f^{(a)}_{\ (b)(c)} A_k^{(b)} \phi^{(c)}
\,.
\label{U1}
\end{equation}

\subsubsection{New instruments for the Lagrangian modeling}

In order to discuss the procedure of the Lagrangian extension, we have to list supplementary tensorial objects, which appear as intrinsic instruments of modeling. For instance, in addition to the spacetime metric $g^{mn}$ one can use the symmetric tensors
\begin{equation}
G_{(a)(b)} U^{(a)m} U^{(b)n} \,, \quad
G_{(a)(b)} \D^m \phi^{(a)} \D^n \phi^{(b)} \,, \quad
G_{(a)(b)} \left(\D^m \phi^{(a)} U^{(b)n} + \D^n \phi^{(a)} U^{(b)m}\right) \,.
\label{U11}
\end{equation}
Similarly, expanding the toolkit for working in a group space, we can use one color vector and four symmetric color tensors.
First, we introduce the multiplet of scalar fields  $\Omega^{(a)}$, which forms the color vector
\begin{equation}
\Omega^{(a)} \equiv \D_m U^{(a)m} =
\nabla_m
U^{(a)m} + g f^{(a)}_{\ (b)(c)} A^{(b)}_m U^{(c)m} \,,
\label{A88}
\end{equation}
the scalar $\Omega$ and the normalized vector $\omega^{(a)}$ in the color space
\begin{equation}
\Omega = \sqrt{\Omega^{(a)} \Omega_{(a)}} \,, \quad  \omega^{(a)} = \frac{\ \Omega^{(a)}}{\Omega} \,.
\label{U77}
\end{equation}
Also, one can use the following color tensors as supplementary instruments of modeling:
\begin{equation}
\omega^{(a)}\omega^{(b)} \,, \quad U^{(a)m}U^{(b)n}g_{mn} \,, \quad \phi^{(a)} \phi^{(b)} \,, \quad \D_m \phi^{(a)} \D_n \phi^{(b)} g^{mn} \,.
\label{U777}
\end{equation}

\subsection{Action functional and variation procedure}

\subsubsection{The structure of the total Lagrangian}

In our work we use the total action functional, which contains three principal parts
\begin{equation}
-S_{(\rm total)} = \int d^4 x \sqrt{{-}g}\left[L_{(\rm CA)} + L_{(\rm YM)} + L_{(\rm PS)}\right] \,.
\label{L1}
\end{equation}
The first part of the Lagrangian $L_{(\rm CA)}$ is marked as the Lagrangian of {\bf C}olor {\bf A}ether
\begin{equation}
L_{(\rm CA)}=\frac{1}{2\kappa}\left[R {+} 2\Lambda {+} \lambda \left(U^m_{(a)}U^{(a)}_m {-}1 \right) {+} {\cal K}^{ijmn}_{(a)(b)}\hat{D}_i U^{(a)}_{m}\hat{D}_j U^{(b)}_n \right] \,.
\label{L2}
\end{equation}
It describes the contributions of the gravity field in the Einstein-Hilbert scheme, and of the multiplet of vector fields. As usual, $R$ is the Ricci scalar, $\Lambda$ is the cosmological constant, $\kappa = 8 \pi G$ is the Einstein constant ($c{=}1$). The Lagrange multiplier $\lambda$ introduces the term, which guarantees that the supernormalization condition exists, $ g_{mn} G_{(a)(b)}U^{(a)m}U^{(a)n} {=} 1$, as the generalization of the relationship $g_{mn} U^m U^n {=}1$ in the canonic theory of the dynamic aether. The object ${\cal K}^{ijmn}_{(a)(b)}$ is a SU(N) symmetric generalization of the Jacobson constitutive tensor. The second term describes the contribution of the {\bf Y}ang - {\bf M}ills field
\begin{equation}
L_{(\rm YM)} = \frac{1}{4} {\cal C}^{mnls}_{(a)(b)} F^{(a)}_{mn} F_{ls}^{(b)} \,,
\label{L3}
\end{equation}
and ${\cal C}^{mnls}_{(a)(b)}$ is the SU(N) symmetric extension of the Tamm constitutive tensor \cite{HO}.
The third term describes the contribution of the {\bf P}seudo {\bf S}calar field
\begin{equation}
L_{(\rm PS)} = \frac12 \Psi^2_0  \left[V(\Phi) {-} {\cal K}^{sl}_{(a)(b)}\D_s \phi^{(a)} \D_l \phi^{(b)} \right] \,.
\label{L4}
\end{equation}
The object ${\cal K}^{sl}_{(a)(b)}$ is the SU(N) symmetric extension of the constitutive tensor, which forms the kinetic term of the pseudoscalar fields in the Lagrangian. The potential $V(\Phi)$ is considered to be the function of the argument  $\Phi {=} \sqrt{\phi^{(a)} \phi^{(b)} G_{(a)(b)}}$; it is chosen to be of the periodic form
\begin{equation}
V(\Phi) = 2 m^2_A \left(1-\cos{\Phi} \right) \,.
\label{14}
\end{equation}
For small values $\Phi$ the potential (\ref{14}) converts into the standard quadratic one $V(\Phi) \to m^2_A \Phi^2$, thus, the parameter $m_A$ has the dimensionality of mass.
The constant $\Psi_0$ is reciprocal to the coupling constant of the axion-gluon interactions $g_{AG}$, i.e.,  $g_{AG}= \frac{1}{\Psi_0}$.

\subsubsection{The structure of constitutive tensors}

({\it i}) {\it Generalized Jacobson's tensor ${\cal K}^{ijmn}_{(a)(b)}$}

In the case, when the symmetry of the system relates to the U(1) group, and when there is only one unit four-vector $U^j$, the canonic Jacobson's constitutive tensor
\begin{equation}
K^{abmn} = C_1 g^{ab} g^{mn} {+} C_2 g^{am}g^{bn}
{+} C_3 g^{an}g^{bm} + C_{4} U^{a} U^{b} g^{mn}
\label{CT1}
\end{equation}
has four phenomenological parameters.
According to \cite{color1} for the case with SU(N) symmetry the decomposition of the color analog of Jacobson's tensor contains 41 elements with corresponding coupling constants. In this work, we focus on the constitutive tensor with two additional coupling constants $\alpha_1$ and $\alpha_2$, which introduce the quadratic combination of the pseudoscalar fields:
\begin{equation}
{\cal K}^{ijmn}_{(a)(b)} =    \left[G_{(a)(b)} {+} \alpha_1 \phi_{(a)} \phi_{(b)} \right] \left[C_1 g^{ij} g^{mn} {+} C_2 g^{im}g^{jn}
{+} C_3 g^{in}g^{jm} \right] +
\label{CT2}
\end{equation}
$$
+ C_{4} \left[U^{i}_{(a)} U^{j}_{(b)} g^{mn} {+} U^{m}_{(b)} U^{n}_{(a)} g^{ij} \right]\left[1{+} \alpha_2 \phi^{(h)}\phi_{(h)}\right] \,.
$$
One can see that  the multiplier $\left[G_{(a)(b)} {+} \alpha_1 \phi_{(a)} \phi_{(b)} \right]$ plays here the role of effective metric in the group space similarly to the effective metric on the spacetime platform (see, e.g., \cite{EM1,EM2,EM3,EM4,EM5,EM6} for terminology, definitions and history details). As for the term $\left[1{+} \alpha_2 \phi^{(h)}\phi_{(h)}\right]$, it modifies in fact the Jacobson coupling constant $C_4$.

\vspace{5mm}

({\it ii}) {\it Generalized color Tamm's tensor ${\cal C}^{ijmn}_{(a)(b)}$}

In the model with U(1) symmetry, the Tamm constitutive tensor for the axionic vacuum has the form \cite{HO}
\begin{equation}
C^{jkmn} = \frac12 \left(g^{im}g^{jn}{-} g^{in}g^{jm} \right) + \frac12 \phi \epsilon^{jkmn} \,.
\label{CT3}
\end{equation}
When the pseudoscalar field is absent, but the multiplet of color four-vectors is present, according to the results of the work \cite{color1}, the decomposition of the color analog of the Tamm tensor contains 12 new coupling constants. In this work we focus on the truncated model with the tensor
\begin{equation}
{\cal C}^{ijmn}_{(a)(b)} = \frac12 \left(g^{im}g^{jn}{-} g^{in}g^{jm} \right)  \left[G_{(a)(b)} {+} \beta_1 \phi_{(a)} \phi_{(b)} {+} \beta_2 \omega_{(a)}\omega_{(b)}\right] +
\label{CT4}
\end{equation}
$$
+ \frac12 \epsilon^{ijmn} \omega^{(c)}\phi^{(d)} \left[G_{(a)(b)} G_{(c)(d)} + \beta_3 \left(G_{(a)(c)} G_{(b)(d)} + G_{(a)(d)} G_{(b)(c)} \right) \right] \,,
$$
which has only three additional coupling constants, $\beta_1$, $\beta_2$, $\beta_3$. In order to clarify the sense of these new parameters, one can calculate the color tensors, which are the generalization of canonic permittivity tensors. If $V^j$ is the velocity four-vector of an observer,
any generalized Tamm's tensor can be decomposed as follows (see, e.g., \cite{HO,EM6} for details):
$$
C^{ikmn}_{(a)(b)} = \frac12 \left[
\varepsilon^{im}_{(a)(b)} V^k V^n - \varepsilon^{in}_{(a)(b)} V^k V^m +
\varepsilon^{kn}_{(a)(b)} V^i V^m - \varepsilon^{km}_{(a)(b)} V^i V^n \right]-
\frac12
\eta^{ikl}(\mu^{-1})_{ls (a)(b)}  \eta^{mns}  -
$$
\begin{equation}
-\frac12 \left[\eta^{ikl}(V^m \nu_{l \ (a)(b) }^{\ n} - V^n \nu_{l
\ (a)(b) }^{\ m}) + \eta^{lmn}(V^i \nu_{l \ (a)(b)}^{\ k} - V^k
\nu_{l \ (a)(b)}^{\ i} ) \right] \,,
\label{Cdecomposition}
\end{equation}
where the tensors of color permittivity, $\varepsilon^{im}_{(a)(b)}$,
color impermeability, $(\mu^{-1})^{pq}_{(a)(b)}$ and color
cross-effect pseudotensors, $\nu_{(a)(b)}^{p m}$ are defined as
\begin{equation}
\varepsilon^{im}_{(a)(b)} {=} 2 {\cal C}^{ikmn}_{(a)(b)} V_k V_n
\,, \quad (\mu^{-1})^{pq}_{(a)(b)} {=} {-} \frac{1}{2} \eta^p_{\
ik} {\cal C}^{ikmn}_{(a)(b)} \eta^{ \ \ \ q}_{mn} \,, \quad
\nu_{(a)(b)}^{p m} {=} \eta^p_{\ ik} {\cal C}^{ikmn}_{(a)(b)} V_n
\,, \label{nu}
\end{equation}
with $\eta^{ikm} \equiv \epsilon^{ikmn} V_n$. For the Tamm tensor (\ref{CT4}) these quantities have the form
\begin{equation}
\varepsilon^{im}_{(a)(b)} = \Delta^{im} \varepsilon_{(a)(b)}
\,, \quad (\mu^{-1})^{pq}_{(a)(b)} {=} \Delta^{pq} (\mu^{-1})_{(a)(b)} \,, \quad
\nu_{(a)(b)}^{p m} {=} \Delta^{pm} \nu_{(a)(b)} \,,
 \label{nu2}
\end{equation}
where $\Delta^{pm} = g^{pm}-V^p V^m$ is the projector, and
\begin{equation}
\varepsilon_{(a)(b)} = G_{(a)(b)} {+} \beta_1 \phi_{(a)} \phi_{(b)} {+} \beta_2 \omega_{(a)}\omega_{(b)} = (\mu^{-1})_{(a)(b)} \,,
\label{nu3}
\end{equation}
\begin{equation}
\nu_{(a)(b)} = - \left[G_{(a)(b)} \omega_{(d)}\phi^{(d)} + \beta_3 \left( \omega_{(a)}\phi_{(b)} + \omega_{(b)}\phi_{(a)}  \right) \right] \,.
\label{nu4}
\end{equation}
This means that the parameters $\beta_1$ and $\beta_2$ mark color modifications of the scalars of permittivities (\ref{nu3}), while the parameter $\beta_3$ introduces color modifications of the canonic cross-effect pseudotensor $\nu^{pm}=-\phi \Delta^{pm}$ \cite{HO}. One can see that the color vectors $\omega^{(a)}$ and $\phi^{(a)}$ create bi-axial anisotropy of the color susceptibilities in the group space.

\vspace{3mm}

({\it iii}) Generalized tensor ${\cal K}^{sl}_{(a)(b)}$

In the canonic model only the metric tensor $g^{sl}$ entered the kinetic part of the Lagrangian of the pseudoscalar field. Based on the analogy with the acoustic metrics (see, e.g., \cite{EM3,EM4,EM5}), in \cite{BS} this tensor was extended along the line of reconstruction of effective spacetime metric; so that the tensor $g^{sl} {+} {\cal A} U^s U^l$ appeared. Now we introduce the extended  color constitutive tensor
\begin{equation}
{\cal K}^{sl}_{(a)(b)}= \left(g^{sl} + \gamma_1 U^s_{(h)}U^{l(h)} \right)  \left(G_{(a)(b)} + \gamma_2 U^j_{(a)}U_{j(b)} + \gamma_3 \omega_{(a)} \omega_{(b)}  \right) +
\label{nu44}
\end{equation}
$$
+\gamma_4 \left(U^s_{(a)}U^l_{(b)}{+} U^s_{(b)}U^l_{(a)} \right)+
\gamma_5 G_{(a)(b)} U^s_{(c)}U^l_{(d)}\omega^{(c)}\omega^ {(d)} \,.
$$
This constitutive tensor contains five new coupling constants $\gamma_1$,...,$\gamma_5$. The parameter $\gamma_1$ plays the role of a kernel of the effective (color-acoustic) spacetime metric. The parameters $\gamma_2$ and $\gamma_3$ are connected with effective metric of a new type in the group space. The parameters $\gamma_4$ and $\gamma_5$ have no analogs, they appear in the context of theory of color aether.

\subsection{Master equations}

\subsubsection{Master equations for the Yang-Mills fields}

Variation of the total action functional (\ref{L1}) with respect to the potentials $A^{(a)}_i$ gives $4(N^2-1)$ extended Yang-Mills equations
\begin{equation}\label{Col1}
\hat{D}_k \left[{\cal C}^{ikls}_{(a)(b)}F_{ls}^{(b)} \right] = \Gamma^i_{(a)}\,,
\end{equation}
where the color current $\Gamma^i_{(a)}$ contains two principal parts, $\Gamma^i_{(a)} = \Gamma^i_{(a)(\rm U)} {+} \Gamma^i_{(a)(\rm PS)}$. The first part
\begin{equation}\label{Cbb1}
 \Gamma^i_{(a)(\rm U)} =    \frac{g}{\kappa} f^{(d)}_{\ (c)(a)}   U^{(c)}_j  {\cal K}^{imjn}_{(d)(b)} \hat{D}_m U^{(b)}_n  +
\frac{g}{4\Omega}f^{(d)}_{\ (c)(a)}  U^{i(c)} \left(\delta^{(q)}_{(d)}-\omega^{(q)}\omega_{(d)} \right) \times
\end{equation}
$$
\times
\left\{ \beta_2 F^{(p)}_{mn} F^{mn(b)} \omega_{(b)} G_{(p)(q)} {+} F^{*(p)}_{mn}F^{mn(b)} \phi^{(f)} \left[G_{(p)(b)}G_{(q)(f)} {+} \beta_3 \left(G_{(p)(q)}G_{(b)(f)} {+} G_{(p)(f)}G_{(b)(q)} \right) \right]\right\}
$$
is the contribution of the multiplet of vector fields; it appears since the covariant derivative $\hat{D}_k U^{(a)j}$ (\ref{DU}) contains the potential $A^{(b)}_j$.
The second part
\begin{equation}\label{Cbb19}
 \Gamma^i_{(a)(\rm PS)} = - g   f^{(d)}_{\ (c)(a)} \Psi^2_0 \phi^{(c)}  {\cal K}^{im}_{(d)(b)} \hat{D}_m \phi^{(b)} - \Psi^2_0 \frac{g}{\Omega}f^{(d)}_{\ (c)(a)}  U^{i(c)} \left(\delta^{(q)}_{(d)}-\omega^{(q)}\omega_{(d)} \right) \times
\end{equation}
$$
\times \D_s \phi^{(p)} \D_l \phi^{(b)} \left[ \gamma_3 \omega_{(b)} G_{(p)(q)} \left(g^{sl} + \gamma_1 U^s_{(h)}U^{l(h)} \right)
+ \gamma_5 G_{(p)(b)} U^s_{(q)} U^l_{(h)} \omega^{(h)}\right]
$$
is, respectively, the contribution of the multiplet of pseudoscalar fields; it is proportional to the parameter $\Psi^2_0$.
In this variation procedure we took into account the following detail
\begin{equation}\label{Cbb2}
\delta \omega^{(c)} = \frac{1}{\Omega} \left[\delta^{(c)}_{(d)}- \omega^{(c)}\omega_{(d)} \right] \delta \Omega^{(d)}
=
\frac{g}{\Omega} f^{(d)}_{\ \ (a)(h)} U^{(h)j} \left[\delta^{(c)}_{(d)}- \omega^{(c)}\omega_{(d)} \right] \delta A^{(a)}_j \,.
\end{equation}

\subsubsection{Master equations for the pseudoscalar fields $\phi^{(a)}$}

Variation of the total action functional with respect to the pseudoscalar fields yields $N^2{-}1$ equations, which can be written in the following convenient form:
\begin{equation}\label{Cbb22}
\D_j \left[{\cal K}^{jl}_{(h)(b)} \D_l \phi^{(b)} \right] + \frac{\phi_{(h)}}{2\Phi} V^{\prime}(\Phi)
=  \frac{1}{\Psi^2_0 }\left[ {\cal J}^{(0)}_{(h)} + {\cal J}^{(1)}_{(h)} \right]\,.
\end{equation}
The first source term in the right-hand side of these equations
\begin{equation}\label{Cbb23}
{\cal J}^{(0)}_{(h)} = - \frac14 F^{(a)*}_{mn} F_{(a)}^{mn} \omega_{(h)} -
\frac12 \beta_3 \omega_{(a)} F^{mn}_{(h)}F_{mn}^{*(a)}
\end{equation}
is of zero order in the pseudoscalar field; it is formed using the Yang-Mills field strength $F_{(a)}^{mn}$ in combination with the multiplier $\omega_{(h)}$, originated from the covariant divergence of the vector field multiplet.
The second term is linear in $\phi_{(b)}$; it does not contain $F_{(a)}^{mn}$, but includes the gauge potential $A^{(a)}_j$, which is the element of the covariant derivative of the vector field $\D_k U^{(h)m}$:
\begin{equation}\label{Cbb24}
{\cal J}^{(1)}_{(h)} = - \frac{\phi_{(b)}}{\kappa} \left\{\alpha_1 \left[C_1 \left(\D^k U_{(h)m}\right)\left(\D_k U^{(b)m}\right) + C_2 \left(\D^m U_{(h)m}\right)\left(\D_n U^{(b)n}\right) + \right. \right.
\end{equation}
$$
\left. \left.+C_3 \left(\D^n U_{(h)m}\right)\left(\D^m U^{(b)}_{n}\right)\right]
\right\} +
$$
$$
+\alpha_2 C_4 \phi_{(h)} \left[\left(U^i_{(a)}\D_i U^{(a)}_{m}\right) \left(U^j_{(b)}\D_j U^{m(b)}\right) + U_{m (b)} U_{n(a)} \D^k U^{m(a)}\D_k U^{n(b)} \right] \,.
$$
We would like to emphasize, that there is the following scheme of the appearance of the pseudoscalar particles. Let us imagine, that the multiplet of vector fields exists; then the corresponding Yang-Mills field appears in order to support the SU(N) symmetry of the system. The Yang-Mills field becomes the source of the pseudoscalar field, since the equation (\ref{Cbb22}) with nonvanishing source  (\ref{Cbb23}) does not admit the solution $\phi^{(a)}=0$.

\subsubsection{Master equations for the vector fields $U^k_{(a)}$}

Variation of the total action functional with respect to the Lagrange multiplier $\lambda$ gives the normalization condition (\ref{4represent}).
Variation with respect to the vector fields $U^j_{(a)}$ gives $4(N^2-1)$ balance equations
\begin{equation}
\D_i {\cal J}^{ij}_{(a)}
 = \lambda \ U^j_{(a)}  +  {\cal I}^{j}_{(a)} \,,
\label{CU1}
\end{equation}
where the color object  ${\cal J}^{ij}_{(a)}$ and the Lagrange multiplier $\lambda$  are introduced in analogy with the canonic aether theory \cite{J1}:
\begin{equation}
{\cal J}^{ij}_{(a)} = {\cal K}^{imjn}_{(a)(b)} \hat{D}_m U^{(b)}_n \,,
\label{CU11}
\end{equation}
\begin{equation}
\lambda =  U_j^{(a)} \left[\D_i {\cal J}^{ij}_{(a)}
- {\cal I}^j_{(a)} \right]  \,.
\label{CU4}
\end{equation}
We took into account that
\begin{equation}
\delta \omega^{(c)} = \frac{1}{\Omega} \left[\delta^{(c)}_{(d)}- \omega^{(c)}\omega_{(d)} \right] \delta \Omega^{(d)}
=
\frac{1}{\Omega} \left[\delta^{(c)}_{(d)}- \omega^{(c)}\omega_{(d)} \right] \D^j \delta U^{(d)}_j \,.
\label{CU109}
\end{equation}
As for the source four-vectors ${\cal I}^j_{(a)}$, they can be written as a sum of three terms
\begin{equation}
{\cal I}^j_{(a)} = {\cal I}^j_{(a)(\rm A)} + {\cal I}^j_{(a)(YM)} + {\cal I}^j_{(a)(\rm PS)} \,.
\label{CU44}
\end{equation}
The first term
\begin{equation}
{\cal I}^j_{(a)(\rm A)} =
C_4 \left(1+ \alpha_2 \phi^{(d)}\phi_{(d)} \right) \left[\hat{D}^j U^m_{(a)}  U^n_{(b)}  \hat{D}_n U_m^{(b)} {+} \hat{D}^k U^j_{(b)}  U_n^{(b)}  \hat{D}_k U^n_{(a)} \right]
\label{CU45}
\end{equation}
is the direct SU(N) generalization of the canonic aetheric current in \cite{J1}.
The second term
\begin{equation}
{\cal I}^j_{(a)(\rm YM)} = - \frac14 \kappa \D^j \left\{ \frac{1}{\Omega}\left[\delta^{(c)}_{(a)} - \omega^{(c)}\omega_{(a)} \right] \times \right.
\label{CU55}
\end{equation}
$$
\left. \times \left[ 2\beta_2 \omega_{(b)} F^{mn}_{(c)} F^{(b)}_{mn}  +
\beta_3 \phi^{(h)} \left(F_{mn(c)} F^{mn*}_{(h)} {+} F_{mn(h)} F^{mn*}_{(c)} \right)
+ \phi_{(c)}F_{mn}^{(d)} F^{mn*}_{(d)}\right] \right\}
$$
is quadratic in the Yang-Mills field strength. The third term
\begin{equation}
{\cal I}^j_{(a)(\rm PS)} = {-}\kappa \Psi^2_0 \gamma_1 U^s_{(a)}\left[G_{(d)(b)} {+} \gamma_2 U^m_{(d)} U_m^{(d)} {+} \gamma_3 \omega_{(d)} \omega_{(b)}\right] (\D_s \phi^{(d)}) (\D^j \phi^{(b)})-
\label{CU56}
\end{equation}
$$
-\kappa \Psi^2_0 \gamma_2 \left[U^j_{(b)} \left(g^{sl}+ \gamma_1 U^s_{(h)} U^{l(h)} \right)\D_l \phi^{(b)} \D_s \phi_{(a)}  \right] -
$$
$$
- \kappa \Psi^2_0 \gamma_4 U^{l(b)} \left[\D_l \phi_{(b)} \D^j \phi_{(a)} + \D_l \phi_{(a)} \D^j \phi_{(b)} \right]
- \kappa \Psi^2_0 \gamma_5 U^s_{(c)} \omega^{(c)} \omega_{(a)}\D^j \phi^{(d)}  \D_s \phi_{(d)} +
$$
$$
{+}\kappa \Psi^2_0 \D^j \left\{\frac{1}{\Omega} \left[\delta^{(c)}_{(a)}{-}\omega^{(c)}\omega_{(a)} \right] \times \right.
$$
$$
\left.\times \left[\gamma_3\left(g^{sl}{+}\gamma_1U^s_{(h)}U^{l(h)} \right)\D_s \phi_{(c)} \D_l \phi^{(b)}\omega_{(b)}
{+}\right.
\left. \gamma_5 U^s_{(c)} U^l_{(d)}\omega^{(d)} \D_s \phi_{(b)} \D_l \phi^{(b)}\right]
\right\} \,.
$$
contains quadratic combinations of the covariant derivatives of the pseudoscalar fields.

\subsubsection{Master equations for the gravitational field}

Variation procedure with respect to the spacetime metric gives the set of ten equations for the gravitational field
\begin{equation}
 R_{pq} {-} \frac{1}{2} R  g_{pq} =  \Lambda g_{pq} + \lambda U^{(a)}_p  U_{(a)q}  +
 T^{(\rm CA)}_{pq} + \kappa T^{(\rm YM)}_{pq} + \kappa T^{(\rm PS)}_{pq} \,.
\label{CE1}
\end{equation}
The stress-energy tensor of the color vector fields is of the following form:
\begin{equation}
T^{(\rm CA)}_{pq} = \frac12 g_{pq} {\cal K}^{ijmn}_{(a)(b)}  \hat{D}_i U^{(a)}_m \hat{D}_j U^{(b)}_n+
\label{CE2}
\end{equation}
$$
{+} G_{(a)(b)}\D^m \left[U^{(b)}_{(p}{\cal J}_{q)m}^{(a)} {-}
{\cal J}_{m(p}^{(a)}U^{(b)}_{q)} {-}
{\cal J}_{(pq)}^{(a)} U^{(b)}_m \right]+
$$
$$
+  C_1[G_{(a)(b)}+\alpha_1\phi_{(a)}\phi_{(b)}]\times[\hat{D}_m U_p^{(a)} \hat{D}^m U_q^{(b)} - \hat{D}_p U^{m(a)} \hat{D}_q U_m^{(b)}] +
$$
$$
+ C_4 [1+\alpha_2\phi^{(h)}\phi_{(h)}] \left\{ U^m_{(a)} \hat{D}_m U^{(a)}_p
U^n_{(b)} \hat{D}_n U^{(b)}_q - U_{m(b)} U_{n(a)} \hat{D}_p U^{m(a)} \hat{D}_q U^{n(b)} \right\} \,.
$$
The parentheses $(pq)$ denote the symmetrization with respect to the coordinate indices $p$ and $q$.

The symbol $T^{(\rm YM)}_{pq}$ refers to the stress-energy tensor of the Yang-Mills field
\begin{equation}
T^{(\rm YM)}_{pq} =
\left[G_{(a)(b)} {+} \beta_1 \phi_{(a)} \phi_{(b)} {+} \beta_2 \omega_{(a)} \omega_{(b)} \right] \left[\frac14 g_{pq} F^{(a)}_{mn} F^{mn (b)} {-} F^{(a)}_{pn} F_q^{(b)n} \right] -
\label{CEH1}
\end{equation}
$$
{-} \frac12 g_{pq}\beta_2 \nabla_j \left\{\frac{\omega_{(b)}}{\Omega} U^{j}_{(c)} F^{(a)}_{mn} F^{(b)mn} \left[\delta_{(a)}^{(c)} {-}\omega_{(a)}\omega^{(c)} \right] \right\} -
$$
$$
- \frac14 g_{pq}\left[G_{(a)(b)} G_{(c)(d)}{+} \beta_3 \left( G_{(a)(c)} G_{(b)(d)} {+} G_{(a)(d)} G_{(b)(c)}\right) \right] \times
$$
$$
\times  \nabla_j \left\{U^{(h)j} F^{(a)}_{mn} F^{*(b)mn} \frac{\phi^{(d)}}{\Omega} \left[\delta^{(c)}_{(h)} {-}\omega^{(c)}\omega_{(h)} \right] \right\} \,.
$$
The last term describes the effective stress-energy tensor of the multiplet of pseudoscalar fields
\begin{equation}
T^{(\rm PS)}_{pq} {=}  \frac12 g_{pq} \Psi^2_0 \left[V(\phi) {-} {\cal K}^{ls}_{(a)(b)}\hat{D}_l \phi^{(a)}\hat{D}_s \phi^{(b)} \right]
{+}
\label{phi179}
\end{equation}
$$
+\Psi^2_0 \hat{D}_p \phi^{(a)}\hat{D}_q \phi^{(b)}\left(G_{(a)(b)} {+} \gamma_2 U^j_{(a)}U_{j(b)} {+} \gamma_3 \omega_{(a)} \omega_{(b)}  \right) {+}
$$
$$
+ \Psi^2_0 \gamma_2 U_{p(a)}U_{q(b)}\left(g^{sl}{+} \gamma_1 U^s_{(h)}U^{l(h)} \right) \hat{D}_l \phi^{(a)} \hat{D}_s \phi^{(b)} +
$$
$$
+ \Psi^2_0 g_{pq} \nabla_n \left\{ \frac{\omega^{(b)}}{\Omega}U^{(d)n}\left[\delta^{(c)}_{(d)}{-}\omega^{(c)}\omega_{(d)} \right] \times \right.
$$
$$
\left. \times \left[\gamma_3 \left(g^{sl}+ \gamma_1 U^{s}_{(h)}U^{l(h)} \right)\hat{D}_s \phi_{(c)} \hat{D}_l \phi_{(b)} + \gamma_5 U^s_{(c)}U^l_{(b)} \hat{D}_s \phi_{(h)} \hat{D}_l \phi^{(h)}
\right]\right\} \,.
$$
We took into account that
\begin{equation}
\delta \omega^{(c)} = \frac{1}{\Omega} \left[\delta^{(c)}_{(d)}- \omega^{(c)}\omega_{(d)} \right] \delta \Omega^{(d)}
=
- \frac{1}{2\Omega} U^{s(d)} g_{pq} \left[\delta^{(c)}_{(d)}- \omega^{(c)}\omega_{(d)} \right] \nabla_s \delta g^{pq} \,.
\label{phi177}
\end{equation}

\vspace{3mm}

{\it REMARK 1}

We derived the complete self-consistent set of coupled master equations for the SU(N) symmetric  system containing four interacting fields: the vector, pseudoscalar, gauge and gravitational fields. Generally, we deal with the system of $9(N^2-1)+10$ coupled equations. Also, there are two sets of equations, which formulate the compatibility conditions. The first one is the differential consequence of the equations (\ref{Col1}) and it has the form
\begin{equation}
\hat{D}_j \hat{D}_k \left[C^{kjmn}_{(a)(b)} F_{mn}^{(b)} \right] \equiv 0   \Rightarrow \hat{D}_j \Gamma^j_{(a)} = 0 \,.
 \label{REM1}
\end{equation}
The second compatibility conditions is the consequence of the Bianchi identity; it has the form
\begin{equation}
\nabla^q \left[R_{pq}-\frac12 g_{pq}R \right] \equiv 0   \Rightarrow \nabla^q \left[\lambda U^{(a)}_p  U_{(a)q}  +
 T^{(\rm CA)}_{pq} + \kappa T^{(\rm YM)}_{pq} + \kappa T^{(\rm PS)}_{pq} \right] = 0 \,.
 \label{REM2}
\end{equation}
It is important to emphasize that the compatibility conditions (\ref{REM1}) and (\ref{REM2}) are performed automatically  on the solutions to the master equations.

\section{Decay of the multi-axionic color aether}

\subsection{Concept of parallel fields}

In classical electrodynamics the so-called electromagnetic polarization tensor $\rho_{\alpha \beta}=\frac{<E_{\alpha} E_{\beta}>}{<E_{\gamma} E^{\gamma}>}$ is used, which is based on the averaged product of the three-dimensional electric field $E_{\alpha}$ (see, e.g., \cite{LL2}). This polarization tensor has the unit trace. When all the eigenvalues of this tensor coincide and are equal to $\frac13$, the configuration of the electromagnetic waves  can be indicated by the term natural light. When one eigenvalue is equal to one, and other two are equal to zero, the light is indicated as polarized. Based on this analogy, we use the term polarized in the context of multi-axionic color aether as follows.

The multiplet of vector fields  can be characterized by the symmetric aetheric color polarization tensor with the unit trace
\begin{equation}
H^{(\rm U)}_{(a)(b)} = \frac{g_{mn} U^m_{(a)} U^n_{(b)}}{g_{ls}G^{(c)(d)} U^l_{(c)} U^s_{(d)}} = U^m_{(a)} U_{(b)m} \,, \quad G^{(a)(b)} H^{(\rm U)}_{(a)(b)} =1 \,.
\label{pp1}
\end{equation}
This tensor has $N^2{-}1$ eigenvectors $q^{[j]}_{(a)}$ and $N^2{-}1$ corresponding eigenvalues $\lambda^{[j]}_{\rm U}$ (the index $[j]$ enumerates the eigenvalues).
Similarly, one can introduce the gauge color polarization tensor
\begin{equation}
H^{(\rm YM)}_{(a)(b)} = \frac{g_{mn} A^m_{(a)} A^n_{(b)}}{g_{mn} G^{(a)(b)}A^m_{(a)} A^n_{(b)}} \,, \quad G^{(a)(b)} H^{(\rm YM)}_{(a)(b)} =1 \,,
\label{pp2}
\end{equation}
with  $N^2{-}1$ eigenvectors $Q^{[j]}_{(a)}$ and $N^2{-}1$ eigenvalues $\lambda^{[j]}_{\rm YM}$. Also, we can deal with  the pseudoscalar color polarization tensor
\begin{equation}
H^{(\rm PS)}_{(a)(b)} = \frac{g^{mn}\D_m \phi_{(a)} \D_n \phi_{(b)}}{g^{mn}G^{(a)(b)}\D_m \phi_{(a)} \D_n \phi_{(b)}} \,, \quad G^{(a)(b)} H^{(\rm PS)}_{(a)(b)} =1
\label{pp3}
\end{equation}
with  $N^2{-}1$ eigenvectors ${\cal Q}^{[j]}_{(a)}$ and $N^2{-}1$ eigenvalues $\lambda^{[j]}_{\rm PS}$.

We consider the color aether to be the polarized one, when
\begin{equation}
U^{(a)m} =  q^{(a)} U^m \,, \quad G_{(a)(b)} q^{(a)} q^{(b)} =1 \,, \quad g_{mn}U^m U^n=1 \,.
\label{pp4}
\end{equation}
Respectively, we obtain the degenerated case, when the obtained aetheric color tensor
\begin{equation}
H^{(\rm U)}_{(a)(b)} = q_{(a)} q_{(b)}
\label{pp5}
\end{equation}
has the eigenvector  $q^{(a)}$ with unit eigenvalue $\lambda^{[1]}_{\rm U}= 1$. In other words, all the vectors of the multiplet $\left\{U^{(a)m} \right\}$ form an one-dimensional subspace in the group space, and thus they are parallel as the vectors in the group space.

Similarly, we consider the gauge field to be polarized, when $A^{(a)}_k = Q^{(a)} A_k$, so that  the gauge color polarization tensor is
\begin{equation}
H^{(\rm YM)}_{(a)(b)} = Q_{(a)} Q_{(b)} \,, \quad G_{(a)(b)}Q^{(a)} Q^{(b)}=1 \,.
\label{pp6}
\end{equation}
In this sense, the Yang-Mills potentials become parallel in the group space, and this model has been used in many works (see, e.g., \cite{Y,G}).
It is important that in this case the Yang-Mills field strength (\ref{46Fmn}) becomes quasi-Abelian due to the skew-symmetric group constants
\begin{equation}
F^{(a)}_{mn} = Q^{(a)} \left(\nabla_m A_m - \nabla_n A_m \right) = Q^{(a)} F_{mn} \,.
\label{pp7}
\end{equation}
Finally, the multiplet of pseudoscalar fields is indicated as the polarized one, when
\begin{equation}
\phi^{(a)} = {\cal Q}^{(a)} \phi  \Rightarrow H^{(\rm PS)}_{(a)(b)} = {\cal Q}_{(a)}{\cal Q}_{(b)} \,, \quad G_{(a)(b)}{\cal Q}^{(a)}{\cal Q}^{(b)}=1 \,.
\label{pp8}
\end{equation}


{\it REMARK 2}

Based on the definitions given above, and keeping in mind that the gauge field can appear, first, in order to guarantee the SU(N) symmetry of the vector or/and pseudoscalar multiplets, second, as an individual non-Abelian player decoupled from the aether and axionic system, we can distinguish  the following interesting models for further consideration.

I. The multi-axionic color aether is completely non-polarized (the general non-Abelian case).

II. The axionic configuration is polarized (there exists only one pseudoscalar field $\phi$ describing the canonic axion field), the vector fields form a non-Abelian multiplet, and the multiplet of gauge potentials is coupled to the vector one.

III. The color aether is polarized (there exists only one vector field $U^j$ describing the canonic dynamic aether), the pseudoscalar fields form a non-Abelian multiplet, and the multiplet of gauge potentials is coupled to the pseudoscalar one.

IV. The color aether and axionic configuration are polarized, but the gauge field is non-Abelian.

V. The multi-axionic color aether is completely polarized, i.e., the vector fields, the Yang-Mills potentials and the axion field multiplet are self-parallel in the group space, simultaneously, i.e.,
\begin{equation}
U^{(a)j}= q^{(a)} U^j, \quad A_j^{(a)} = Q^{(a)} A_j \,, \quad  \phi^{(a)} = {\cal Q}^{(a)} \phi \,.
\label{pp9}
\end{equation}
In other words there are three one-dimensional subspaces in the group space, which are based on the color vectors $q^{(a)}$, $Q^{(a)}$ and ${\cal Q}^{(a)}$.
In this context there is a number of particular sub-cases:

V.1. $q^{(a)} \neq Q^{(a)} \neq {\cal Q}^{(a)}$;

V.2. $q^{(a)} \neq Q^{(a)} = {\cal Q}^{(a)}$;

V.3. $q^{(a)} =  Q^{(a)} \neq {\cal Q}^{(a)}$;

V.4. $q^{(a)} = {\cal Q}^{(a)} \neq  Q^{(a)}$;

V.5. $q^{(a)} = Q^{(a)} = {\cal Q}^{(a)}$.

Clearly, in the cases V.2. and V.5. we obtain
\begin{equation}
\D_k \phi^{(a)} = {\cal Q}_{(a)} \nabla_k \phi + g f^{(a)}_{(b)(c)} A_k \phi  Q^{(b)} Q^{(c)} = {\cal Q}_{(a)} \nabla_k \phi \,,
\label{pp10}
\end{equation}
and in the cases V.3. and V.5. we see that
\begin{equation}
\D_k U^{j(a)} = q^{(a)} \nabla_k U^j \,.
\label{pp11}
\end{equation}

\subsection{Bianchi-I spacetime platform}

Master equations for the established model are obtained for arbitrary set of phenomenologically introduced coupling constants. But we are interested to apply this theory to the cosmological model with high spacetime symmetry. The homogeneous anisotropic model of the  Bianchi-I type is a typical candidate for this purpose. Clearly, the symmetry requirements associated with our choice impose restrictions on these parameters, and the corresponding reduced model  turns out inevitably to be truncated. Let us consider these restrictions and basic formulas.

\subsubsection{Geometric aspects of the model}

The spacetime in the Bianchi-I model belongs to the class of spatially homogeneous and anisotropic ones \cite{LL2,Bianchi}.
The spacetime metric can be presented in the form
\begin{equation}
ds^2 = dt^2 - a^2(t) {dx^1}^2 - b^2(t) {dx^2}^2 - c^2(t) {dx^3}^2 \,.
\label{BI1}
\end{equation}
The non-vanishing components of the Einstein tensor $G^k_m = R^k_m -\frac12 \delta^k_m R$ are
\begin{equation}
G^0_0 = \frac{\dot{a}}{a} \frac{\dot{b}}{b} + \frac{\dot{a}}{a} \frac{\dot{c}}{c} + \frac{\dot{b}}{b} \frac{\dot{c}}{c} \,, \quad
G^1_1 = \frac{\ddot{b}}{b} +  \frac{\ddot{c}}{c} + \frac{\dot{b}}{b} \frac{\dot{c}}{c} \,,
\label{BI2}
\end{equation}
$$
G^2_2 = \frac{\ddot{a}}{a} +  \frac{\ddot{c}}{c} + \frac{\dot{a}}{a} \frac{\dot{c}}{c} \,, \quad
G^3_3 = \frac{\ddot{b}}{b} +  \frac{\ddot{a}}{a} + \frac{\dot{b}}{b} \frac{\dot{a}}{a} \,.
$$
Hence, the first restriction is that non-diagonal components of the total stress-energy tensor have to be vanishing.
Also, we assume that all the physical fields inherit the spacetime symmetry, and thus also depend on the cosmological time only, $\phi^{(a)}(t)$, $A^{(a)}_m(t)$ and $U^{(a)j}(t)$.

\subsubsection{Coulomb gauge conditions}

We assume that when the Yang-Mills field becomes polarized in the Bianchi-I spacetime, i.e., $A^{(a)}_j=Q^{(a)} A_j$, the Lorentz gauge condition takes place:
\begin{equation}
\nabla_k A^k = 0 \Rightarrow \frac{1}{abc} \partial_k \left[abc A^k \right] = 0 \Rightarrow A^0 = \frac{\rm const}{abc} \,.
\label{BI3}
\end{equation}
We assume that the constant is vanishing, and thus $A_0(t)=0$. Also, we assume that the Landau gauge is fulfilled, if the color aether is polarized, i.e., $U^{(a)j}=q^{(a)}U^j$:
\begin{equation}
U_k A^k =0 \Rightarrow U_{\alpha} A^{\alpha} =0  \,.
\label{BI4}
\end{equation}
Here and below the Greek indices run over three values $\alpha = 1,2,3$. The combination of the Lorentz and Landau conditions is known as the Coulomb gauge, and we can guaranty the Coulomb conditions, in particular, if
\begin{equation}
U_{\alpha}=0 \,, \quad U^j = \delta^j_0 \,, \quad A_0 = 0\,.
\label{BI5}
\end{equation}

\subsubsection{Ansatz concerning the aether velocity four-vector $U^j$}

We assume that the polarized color aether has the velocity four-vector of the form $U^j=\delta^j_0$, and this assumption should not be in conflict with all the master equations of the model. This is the second important restriction on the phenomenological coupling constants. Then,
the covariant derivative of the aether velocity four-vector is of the form
\begin{equation}
U^j = \delta^j_0 \Rightarrow \nabla_k U^m = \delta^m_1 \delta_k^1 \left(\frac{\dot{a}}{a}\right) + \delta^m_2 \delta_k^2 \left(\frac{\dot{b}}{b}\right) + \delta^m_3 \delta_k^3 \left(\frac{\dot{c}}{c}\right) \,.
\label{BI6}
\end{equation}
The tensor $\nabla_k U_m$ contains the symmetric part only, and the following  decomposition exists:
\begin{equation}
\nabla_k U^m = \sigma_{k}^m +\frac13 \Theta \Delta^m_{k} \,,
\label{BI7}
\end{equation}
where
\begin{equation}
\Theta = \nabla_k U^k = \frac{\dot{a}}{a}+\frac{\dot{b}}{b}+\frac{\dot{c}}{c} \,,
\label{BI8}
\end{equation}
the projector is now formed using the aether velocity four-vector
\begin{equation}
 \Delta^k_{m} = \delta^k_{m} - U^k U_m \,,
\label{BI9}
\end{equation}
and the symmetric shear tensor has the following non-vanishing components:
\begin{equation}
\sigma^1_1 = \frac{\dot{a}}{a} - \frac13 \Theta \,, \quad \sigma^2_2 = \frac{\dot{b}}{b} - \frac13  \Theta \,, \quad \sigma^3_3 = \frac{\dot{c}}{c} - \frac13 \Theta \,.
\label{BI10}
\end{equation}
The acceleration four-vector and the vorticity tensor are vanishing
\begin{equation}
{\cal D}U^j = U^k \nabla_k U^j = 0 \,, \quad \omega_{jl} = \Delta^m_j \Delta^n_l \left(\nabla_m U_n - \nabla_n U_m \right)=0 \,.
\label{BI11}
\end{equation}
Also, when we deal with the polarized gauge and vector fields, the Landau gauge $A_k U^k = 0$ provides that
\begin{equation}
\Omega^{(a)} = q^{(a)} \nabla_k U^k + g f^{(a)}_{\ (b)(c)} Q^{b}q^{(c)} A_k U^k = q^{(a)} \Theta \,.\label{BI12}
\end{equation}
This means that $\Omega = \Theta$ and thus $\omega^{(d)}= q^{(d)}$ is constant color vector in the group space.

Finally, we see that for the quasi-Abelian potential $(0, A_1(t), A_2(t), A_3(t)$) only the components $F_{0 \alpha}$ of the quasi-Maxwell tensor are non-vanishing, and thus the pseudo-invariant $F^{*}_{mn}F^{mn}$ is equal to zero. Also, we have to emphasize that the stress-energy tensor of the reduced Yang-Mills field has no non-diagonal components, when only one component of the potential is non-vanishing, say, $A_3(t)\neq 0$ only.

\subsection{Test model: completely polarized system with parallel fields}

As a first step of analysis of the model, we have to check our ansatz using the test model of the Bianchi-I type. We assume that at $t=t_*$ all three interacting fields happened to be spontaneously degenerated, i.e., they become parallel in the group space, and $q^{(a)} {=} Q^{(a)} {=} {\cal Q}^{(a)}$. In this case the following simplifications are valid for the cosmological epoch  $t>t_*$:
\begin{equation}
\hat{D}_k \phi^{(a)} = q^{(a)}U_k \dot{\phi} \,, \quad \hat{D}_k U^{(a)j} = q^{(a)} \nabla_k U^j \,, \quad F^{(a)}_{mn} = q^{(a)} \left(U_m \dot{A}_n - U_n \dot{A}_m  \right) \,, \quad \omega^{(c)} = q^{(c)}\,.
\label{BI13}
\end{equation}
The dot denotes the derivative with respect to cosmological time.

\subsubsection{Behavior of the aetheric vector field}

In the situation described above the Jacobson tensor (\ref{CU11}) is equal to
\begin{equation}
{\cal J}^{ij}_{(a)}= q_{(a)}(1+\alpha_1 \phi^2) C_2 g^{ij} \Theta \,,
\label{BI139}
\end{equation}
and the master equations for the aether velocity four-vector (\ref{CU1}) take the form
\begin{equation}
q_{(a)} U^j \left\{ \frac{d}{dt} \left[(1+\alpha_1 \phi^2) C_2 \Theta \right] -
\lambda + \kappa\Psi^2_0 {\dot{\phi}}^2 \left[\gamma_1  \left(2\gamma_2+\gamma_3 \right)
+\gamma_1+\gamma_2 +2\gamma_4 +\gamma_5 \right]  \right\}=0 \,.
\label{BI14}
\end{equation}
Clearly, $4(N^2-1)$ master equations (\ref{BI14}) convert into one equation for the Lagrange multiplier
\begin{equation}
\lambda =  \frac{d}{dt} \left[(1+\alpha_1 \phi^2) C_2 \Theta \right]
 + \kappa\Psi^2_0 {\dot{\phi}}^2 \left[\gamma_1  \left(2\gamma_2+\gamma_3 \right)
+\gamma_1+\gamma_2 +2\gamma_4 +\gamma_5 \right]
\label{lambda11}
\end{equation}
for $U^j =\delta^j_0$, for arbitrary $q^{(a)}$ and arbitrary set of coupling parameters $\alpha_1$, $\gamma_1$,..., $\gamma_5$. In other words, the Jacobson equations for the aether velocity are solved on the interval $t>t_*$.

\subsubsection{Behavior of the gauge field}

The Yang-Mills equations (\ref{Col1}) for the time interval $t>t_*$ can be simplified as follows.
The induction tensor takes the form
\begin{equation}
{\cal C}^{ijmn}_{(a)(b)} F^{(b)}_{mn} = q_{(a)} \left[F^{ij}  \left(1+ \beta_1 \phi^2  {+} \beta_2 \right)
+ F^{*ij} \phi (1+2\beta_3) \right] \,,
\label{BI16}
\end{equation}
and the generalized color current vanishes $\Gamma^i_{(a)} =0$.
Thus, the reduced Yang-Mills equations can be written as
\begin{equation}
\frac{1}{abc} \frac{d}{dt}\left[abc F^{\nu 0} q_{(a)} \left(1+ \beta_1  \phi^2 {+} \beta_2\right) \right]  = 0 \,.
 \label{BI17}
\end{equation}
Since $q^{(a)}$ is a constant, we can omit this multiplier and obtain three equations for three components of the Yang-Mills potential.
There are two principal cases. First, when $1+ \beta_1  \phi^2 {+} \beta_2 =0$, the components $F^{\nu 0}$ are arbitrary functions of the cosmological time. When $1+ \beta_1  \phi^2 {+} \beta_2 \neq 0$, integration of the equations (\ref{BI17}) gives
\begin{equation}
 F^{\nu 0 } = \frac{\rm const}{abc \left(1+ \beta_1  \phi^2 {+} \beta_2\right)} \,,
 \label{BI18}
\end{equation}
and we can now find the corresponding potentials. Mention should be made that the stress-energy tensor of such Yang-Mills field has no non-diagonal terms, if the potential $A_j$ has only one component, say, $A_3(t)$. Then
\begin{equation}
\dot{A}_{3}(t) = {\rm const}\frac{c}{ab \left(1+ \beta_1  \phi^2 {+} \beta_2\right)} \,,
\label{BI19}
\end{equation}
and thus
\begin{equation}
A_{3}(t) = A_3(t_*) + \dot{A}_{3}(t_*) \int_{t_*}^{t} d\tau  \frac{a(t_*) b(t_*) c(\tau) \left(1+ \beta_1  \phi^2(t_*) {+} \beta_2\right)}{a(\tau) b(\tau) c(t_*) \left(1+ \beta_1  \phi^2(\tau) {+} \beta_2\right)} \,.
\label{BI199}
\end{equation}

\subsubsection{Behavior of the axion field}

In order to reduce the equations for the pseudoscalar fields (\ref{Cbb22}), we have to simplify, first, the generalized constitutive tensor ${\cal K}^{sl}_{(a)(b)}$, which takes the form
\begin{equation}\label{Cbb074}
{\cal K}^{sl}_{(a)(b)}= \left(g^{sl} + \gamma_1 U^sU^{l} \right) \left[G_{(a)(b)} + \left(\gamma_2 + \gamma_3\right) q_{(a)}q_{(b)}  \right] +
\left[\gamma_5 G_{(a)(b)} + 2\gamma_4 q_{(a)}q_{(b)} \right] U^sU^l \,.
\end{equation}
The source term ${\cal J}^{(0)}_{(h)}$ given by (\ref{Cbb23}) happens to be equal to zero, since now the pseudoscalar $F^{*}_{mn} F^{mn}$ is vanishing.
The source term ${\cal J}^{(1)}_{(h)}$ given by (\ref{Cbb24}) contains only the term proportional to the constant $C_2$:
\begin{equation}\label{Cbb274}
{\cal J}^{(1)}_{(h)} = - \frac{\phi}{\kappa} q_{(h)}\alpha_1 C_2 \Theta^2  \,.
\end{equation}
Indeed, for the symmetrical tensor $\nabla_mU_n$ two terms disappear due to the condition $C_1+C_3=0$, appeared after estimation of the velocity of the gravity waves (see, e.g., \cite{PSR17,Dark}). The term proportional to the constant $C_4$ disappears since the acceleration vector ${\cal D}U^j$ vanishes.
The common multiplier $q_{(h)}$ can be omitted, and thus we obtain one equation for the axion field $\phi$
\begin{equation}\label{Cbb922}
\Gamma^* \left(\ddot{\phi} + \Theta \dot{\phi}\right)  + \frac{1}{2}  V^{\prime}(\phi) + \frac{\phi}{\kappa \Psi^2_0} \alpha_1 C_2 \Theta^2  =0 \,,
\end{equation}
where we used the convenient notation
\begin{equation}\label{Cbb923}
\Gamma^* = (1+\gamma_1)(1+ \gamma_2 + \gamma_3) + 2 \gamma_4 + \gamma_5 \,.
\end{equation}
Initial data for (\ref{Cbb922}), i.e., $\phi(t_*)$ and $\dot{\phi}(t_*)$ can be fixed if the solutions to the master equations for the interval $t<t_*$ are found.

\subsubsection{Behavior of the gravitational field}

In order to write the master equations of the gravity field (\ref{CE1}) for the interval of the cosmological time $t>t_*$, we have to use the formula (\ref{lambda11}) for the Lagrange multiplier $\lambda$, the reduced formula
\begin{equation}
T^{p(\rm CA)}_{q} = - \delta^{p}_{q} C_2 \left\{ \frac12 \left(1 {+} \alpha_1 \phi^2 \right) \Theta^2 {-} \frac{d}{dt} \left[(1+\alpha_1 \phi^2) \Theta  \right]\right\}
\label{CE29}
\end{equation}
for the stress-energy tensor of the color aether (\ref{CE2}), the reduced formula
\begin{equation}
T^{p(\rm YM)}_{q} = -\frac{1}{c^2(t)} F^2_{03}
\left(1 + \beta_2 + \beta_1 \phi^2 \right)  \left(\frac12 \delta^{p}_{q} - \delta^{p}_{0}\delta^{0}_{q} - \delta^{p}_{3}\delta^{3}_{q} \right)
\label{CEM19}
\end{equation}
for the stress-energy tensor of the Yang-Mills field (\ref{CEH1}) with
\begin{equation}
F_{03} =
\dot{A}_{3}(t_*) \frac{a(t_*) b(t_*) c(t) \left(1+ \beta_1  \phi^2(t_*) {+} \beta_2\right)}{a(t) b(t) c(t_*) \left(1+ \beta_1  \phi^2(t) {+} \beta_2\right)}
\,,
\label{CEM199}
\end{equation}
and the reduced formula
\begin{equation}
T^{p(\rm PS)}_{q} {=}  \frac12 \delta^{p}_{q} \Psi^2_0 \left\{V(\phi) {-} \left[\left(1 + \gamma_1 \right) \left(1 + \left(\gamma_2 + \gamma_3\right) \right) +
\gamma_5 + 2\gamma_4 \right] {\dot{\phi}}^2  \right\}
{+}
\label{phi979}
\end{equation}
$$
+\Psi^2_0 U^p U_q  {\dot{\phi}}^2 \left[1 {+} \gamma_2 {+} \gamma_3 {+} \gamma_2\left(1{+} \gamma_1 \right)\right]
$$
for the stress-energy tensor of the pseudoscalar field (\ref{phi179}). Then four nontrivial Einstein's equations for the Bianchi-I spacetime model take the form
\begin{equation}
\frac{\dot{a}}{a} \frac{\dot{b}}{b} + \frac{\dot{a}}{a} \frac{\dot{c}}{c} + \frac{\dot{b}}{b} \frac{\dot{c}}{c} - \Lambda =
-\frac12  \left(1 {+} \alpha_1 \phi^2 \right)C_2 \Theta^2 + \frac{1}{2c^2(t)} \kappa \left[1 + \beta_2 + \beta_1 \phi^2 \right]F^2_{03} +
\label{00}
\end{equation}
$$
+ \frac12 \kappa \Psi^2_0 \left\{V(\phi)+
{\dot{\phi}}^2 \left[ (1+\gamma_1)(1+\gamma_3+ 5\gamma_2)+ 2\gamma_4 + \gamma_5 \right]\right\} \,,
$$
\begin{equation}
\frac{\ddot{b}}{b} +  \frac{\ddot{c}}{c} + \frac{\dot{b}}{b} \frac{\dot{c}}{c} - \Lambda =
-\frac12 \left(1 {+} \alpha_1 \phi^2 \right)C_2 \Theta^2 {-} \frac{d}{dt} \left[(1+\alpha_1 \phi^2) C_2 \Theta  \right] +
\label{11}
\end{equation}
$$
-\left[1 + \beta_2 + \beta_1 \phi^2 \right] \frac{1}{2c^2(t)} F^2_{03} +
\frac12 \Psi^2_0 \left\{V(\phi) {-} \left[\left(1 + \gamma_1 \right) \left(1 + \left(\gamma_2 + \gamma_3\right) \right) +
\gamma_5 + 2\gamma_4 \right] {\dot{\phi}}^2  \right\}\,,
$$
\begin{equation}
\frac{\ddot{a}}{a} +  \frac{\ddot{c}}{c} + \frac{\dot{a}}{a} \frac{\dot{c}}{c} - \Lambda = -\frac12 \left(1 {+} \alpha_1 \phi^2 \right)C_2 \Theta^2 {-} \frac{d}{dt} \left[(1+\alpha_1 \phi^2) C_2 \Theta  \right] +
\label{22}
\end{equation}
$$
-\left[1 + \beta_2 + \beta_1 \phi^2 \right] \frac{1}{2c^2(t)} F^2_{03} +
\frac12 \Psi^2_0 \left\{V(\phi) {-} \left[\left(1 + \gamma_1 \right) \left(1 + \left(\gamma_2 + \gamma_3\right) \right) +
\gamma_5 + 2\gamma_4 \right] {\dot{\phi}}^2  \right\} \,,
$$
\begin{equation}
\frac{\ddot{b}}{b} +  \frac{\ddot{a}}{a} + \frac{\dot{b}}{b} \frac{\dot{a}}{a} - \Lambda =
-\frac12 \left(1 {+} \alpha_1 \phi^2 \right)C_2 \Theta^2 {-} \frac{d}{dt} \left[(1+\alpha_1 \phi^2) C_2 \Theta  \right] +
\label{33}
\end{equation}
$$
+\left[1 + \beta_2 + \beta_1 \phi^2 \right] \frac{1}{2c^2(t)} F^2_{03} +
\frac1 2 \Psi^2_0 \left\{V(\phi) {-} \left[\left(1 + \gamma_1 \right) \left(1 + \left(\gamma_2 + \gamma_3\right) \right) +
\gamma_5 + 2\gamma_4 \right] {\dot{\phi}}^2  \right\} \,.
$$

\subsection{Two examples of exact solutions of the truncated model}

\subsubsection{Axion field is in an equilibrium state}

When the axion field is frozen into one of the minima of the potential $V(\phi)= 2m^2_A\left(1 {-} \cos{\phi} \right)$, i.e., when $\phi = 2\pi k$, where $k$ is an integer, we   have to use a truncated model, for which $\alpha_1=0$ (see (\ref{Cbb922})). Also, we assume that
the free parameter $\beta_2$ is chosen so that $1 {+} \beta_2 {+} 4 \beta_1 \pi^2 k^2=0$, and obtain that the contribution of the Yang-Mills field is absent in the gravity field equations. Finally, we assume that the model is of the Locally Invariant type with respect to rotations around the $x^1 O x^2$ axis, i.e., $a(t)=b(t)$. Now we deal with the following three  gravity field equations:
\begin{equation}
-3 H^2 + 2 H \Theta =  \Lambda -\frac12  C_2 \Theta^2 \,,
\label{005}
\end{equation}
\begin{equation}
 \dot{\Theta} - \dot{H} + 3H^2 + \Theta (\Theta - 3 H)= \Lambda -\frac12 C_2 \Theta^2 {-} C_2 \dot{\Theta}\,,
\label{325}
\end{equation}
\begin{equation}
2 \dot{H} + 3H^2 = \Lambda -\frac12 C_2 \Theta^2 {-} C_2 \dot{\Theta}\,,
\label{335}
\end{equation}
where we used the formulas $H(t)= \frac{\dot{a}}{a}$ and $\frac{\dot{c}}{c} = \Theta -2H$.
Typically, one of these three equations is the differential consequence of  two others. Also, one can see another consequence
\begin{equation}
 \dot{\Theta} - 3H = - \Theta (\Theta - 3H)  \Rightarrow \Theta(t) = 3H(t) + {\cal K} \frac{a^2(t_*)c(t_*)}{a^2(t)c(t)}\,, \quad {\cal K} = \Theta(t_*) - 3H(t_*) \,,
\label{395}
\end{equation}
which guarantees that asymptotically $\Theta \to 3H$ or equivalently $\frac{\dot{c}}{c} \to \frac{\dot{a}}{a} $. It is the signal of the Universe isotropisation. From the equations (\ref{005}) and (\ref{395}) we obtain the following solutions:
\begin{equation}
H = - \frac13 {\cal K} \left[\left(\frac{a(t_*)}{a(t)} \right)^2 \left(\frac{c(t_*)}{c(t)} \right) \right]  \pm
\sqrt{H^2_{\infty} + \frac{{\cal K}^2}{9\left(1+\frac32C_2 \right)}  \left[\left(\frac{a(t_*)}{a(t)} \right)^4 \left(\frac{c(t_*)}{c(t)} \right)^2 \right]
} \,,
\label{795}
\end{equation}
\begin{equation}
\Theta = \pm
3 \sqrt{H^2_{\infty} + \frac{{\cal K}^2}{9\left(1+\frac32C_2 \right)}  \left[\left(\frac{a(t_*)}{a(t)} \right)^4 \left(\frac{c(t_*)}{c(t)} \right)^2 \right]
} \,, \quad H_{\infty} = \sqrt{\frac{2\Lambda}{3(2+3C_2)}}\,.
\label{895}
\end{equation}
When the Universe expands as a whole and the so-called unit volume $V(t)=a(t)b(t)c(t)$ increases with time, we have to choose the sign plus in (\ref{795}) and (\ref{895}) providing the scalar $\Theta$ to be positive. The function $H(t)$ is positive in two cases.
First, when $- \frac23 <C_2<0$; second, when $C_2>0$ and $H^2_{\infty} > \frac{C_2{\cal K}^2}{3(2+3C_2)}$. When ${\cal K}=0$, we deal with the special sub-model, for which $\Theta(t) = 3 H(t) = 3 H_{\infty}$, or equivalently,  $\frac{\dot{c}}{c}=\frac{\dot{a}}{a} = H_{\infty}$.

Clearly, one can rewrite the equations under discussion in the form of the two-dimensional autonomous dynamic system
\begin{equation}
\dot{H} = \frac{1}{2(2+3C_2)} \left(2 \Lambda + C_2 \Theta^2 - 6H^2 -6C_2 H \Theta \right)\,,
\label{DS1}
\end{equation}
$$
\dot{\Theta}=
\frac{1}{2(2+3C_2)} \left(6 \Lambda - \Theta^2(4+3C_2) - 18H^2 + 12 H \Theta \right)
\,.
$$
When $\Lambda > 0$ and $1+\frac32 C_2>0$, this dynamic system has only one stationary point with $H=H_{\infty}$, $\Theta=3H_{\infty}$. This stationary point is the stable attracting radial sink.

\subsubsection{Axion field is in a dynamic equilibrium state}

The term dynamic equilibrium means that the solution to the axion field equation (\ref{Cbb922}) is constant, i.e., $\phi = \phi_* = const$, but now this constant solution satisfies equation
\begin{equation}
\frac12 V^{\prime}(\phi_*) + \frac{\phi_*}{\kappa \Psi^2_0} \alpha_1 C_2 \Theta^2 =0  \,,
\label{DS11}
\end{equation}
 or in more detail
\begin{equation}
\sin{\phi_*} = - \frac{\phi_*}{\kappa m^2_A  \Psi^2_0} \alpha_1 C_2 \Theta^2 \,.
\label{DS12}
\end{equation}
Clearly, this solution exists, if $\Theta = {\rm const}$, but the discussion in the previous subsubsection shows that this is possible, if the following three conditions are fulfilled:

1. $1 {+} \beta_2 {+} \beta_1 \phi^2_* =0$, i.e., the gauge field is switched out from the gravity field equations;

2. nontrivial solutions to the equation (\ref{DS12}) exist; for instance, it is possible, when $\alpha_1 C_2 < 0$ and $|\alpha_1 C_2| < \frac{\kappa m^2_A \Psi^2_0}{9H^2_{\infty}}$;

3. $\Theta= 3H = 3 H^*_{\infty}$, i.e., the gravity field equations are satisfied, and the asymptotic value of the Hubble function is
\begin{equation}
H^*_{\infty} = \sqrt{\frac{\Lambda + m^2_A \Psi^2_0 (1-\cos{\phi_*})}{3\left[1+ \frac32 C_2 \left(1 {+} \alpha_1 \phi^2_* \right) \right]}} \,.
\label{HHH}
\end{equation}

\section{Outlook}

In the presented work we formulated the theory of interaction between three SU(N) symmetric multiplets of vector, pseudoscalar and gauge fields coupled by the gravitational field. For short, we indicated the corresponding field configuration as multi-axionic color aether, and
made an attempt to motivate this definition linguistically. The total Lagrangian of the presented theory contains ten new coupling parameters, which are introduced phenomenologically. For this Lagrangian the self-consistent set of coupled master equations for the vector, pseudoscalar, gauge and gravitational fields is derived. Then we formulated the concept of polarization of the multi-axionic color aether, which is based on the idea of alignment of the vector, pseudoscalar and gauge field multiplets in the group space associated with the SU(N) group. This idea is well known for the gauge field (see, e.g., \cite{Y,G}), but we extend it for the vector and pseudoscalar field multiplets.

In fact, this work presents a comprehensive research program, which is now partially realized.
Indeed, in the works \cite{color2,color3} the model with axionic singlet and SU(N) symmetric vector and gauge field multiplets
was studied. In the works \cite{STFI1,STFI2} we focused on the analysis of the evolution of the non-Abelian SU(2) symmetric Yang-Mills field of the magneto-electric type for the case, when the pseudoscalar multiplet is presented by one axionic field, and the color aether is converted into the canonic dynamic aether. In the last section of this paper we studied the test model of the completely polarized multi-axionic color aether.
We hope to present soon the results obtained for the model, in the framework of which the multiplet of pseudoscalar fields is  non-Abelian, but the vector and gauge fields are polarized. Other sub-models, listed above in the REMARK 2, are still awaiting their description.

%
%
%

\end{document}